\documentclass[useAMS,usenatbib]{mn2e}
\usepackage[T1]{fontenc}
\usepackage{graphicx}

\newcommand{\av}[1]{\langle{#1}\rangle}

\title{Weak lensing and the Dyer--Roeder approximation}

\author[K. Bolejko]{K. Bolejko \\
Centre for Gravitational Physics, Department of Quantum Science, Australian National University, Canberra, ACT 0200, Australia}

\begin{document}

\maketitle
\label{firstpage}

\maketitle

\begin{abstract}
The distance-redshift relation plays an important role in cosmology.
In the standard approach to cosmology it is assumed that this relation is the same as in the
homogeneous universe. As the real universe is not homogeneous there
are several methods to calculate the correction. The weak lensing approximation
and the  Dyer--Roeder relation are one of them.
This paper establishes a link between these two approximations. 
It is shown that if the universe is homogeneous with only small,
vanishing after averaging, density fluctuations along the line of sight, then
the distance correction is negligible.
It is also shown that a vanishing 3D average of
density fluctuations does not imply 
that the mean of density fluctuations along the line of sight is zero.
In this case, even within the linear approximation, the distance correction is not negligible.
The modified version of the Dyer--Roeder relation is presented
and it is shown that this modified relation is consistent with the correction obtained within the weak lensing approximation.
The correction to the distance for a source at $z\sim 2$ is of order of a few percent.
Thus, with an increasing precision of cosmological observations 
an accurate estimation of the distance is essential. Otherwise errors 
due to miscalculation the distance can become a major source of systematics.
\end{abstract}


\section{Introduction}

The distance-redshift relation plays an important role in cosmology.
In fact almost all cosmological observations depend,  
either explicitly or implicitly, on this relations.
However, in general case, 
to calculate the distance, knowledge of matter distribution (as well as its evolution) between the observer and the source is needed.
Therefore, an approximate relations are of great use to astronomers.
The simplest approximation assumes homogeneity
of the universe and is based on the Friedmann model.
However, as the real universe is not homogeneous 
a more elaborate relations are needed.
Zel'dovich (1964)  proposed an approximation which takes into account that light propagates 
through emptier rather than denser regions of the universe.
This approximation is now known as the  Dyer--Roeder approximation (Dyer \& Roeder 1972, 1973).
It also assumes homogeneity but allows for a different density 
than in the background model. The density difference is modelled 
by a constant, the so-called smoothness parameter
$\alpha$. As the Dyer--Roeder equation is a differential equation
an approximate analytic solution was presented and discussed by Demia\'nski et al. (2003).
Generalisation to  $\alpha(z)$ was first suggested by Linder (1988)
and the effect of the change of the expansion rate was discussed by  Mattsson (2010).
The Dyer--Roeder relation was tested against cosmological observations
by Santos \& Lima (2006), Santos, Cunha \& Lima (2008), and Yu et al. (2010).

However, it has been argued that the Dyer--Roeder relation
may not properly describe the effect of 
matter clustering (R\"as\"anen 2009; Ellis 2009)
and therefore may not be an appropriate approximation for the
distance-redshift relation in the real universe.
Also, if the smoothness parameter changes with redshift 
the distance formula depends on several free-parameters
whose interpretation can be ambiguous. 
Therefore, although cosmologists are aware that 
not taking into account inhomogeneities introduce
additional systematics, the Dyer--Roeder relation
is not widely used. 
Instead, several alternatives based either on the linear perturbative
scheme (see {Bonvin, Durrer \& Gasparini 2006} and references therein)
or non-linear models (see {Bolejko et al. 2009} and references therein)
have been developed.

This paper explores the connection between 
the Dyer--Roeder approximation and the weak lensing
approximation.
{ It shows that the Dyer--Roeder relation can be modified 
so that it is consistent with the lensing
approximation. The modified version 
contains only 1 free parameter of clear interpretation, i.e.
the mean of density fluctuations along the line of sight $\av{\delta}_{1D}$.}

The structure of this paper is as follows:
Sections \ref{DRa} and \ref{Lena} present the Dyer--Roeder 
and weak lensing approximations respectively,
Sec. \ref{res} presents the comparison
of these two methods, and Sec. \ref{dis} discusses the results.

\section{the Dyer--Roeder approximation}\label{DRa}

The angular diameter distance $D_A$
is given by the following relation (Sachs 1961)
\begin{equation}
\frac{{\rm d^2} D_A}{{\rm d} s^2} = - ( |\sigma|^2 + \frac{1}{2} R_{\alpha
\beta} k^{\alpha} k^{\beta}) D_A, \label{dsr}
\end{equation}
where $\sigma$ is the shear of the light bundle, $k^\alpha$ is a vector tangent to the light ray,
$R_{\alpha \beta}$ is the Ricci tensor, and $R_{\alpha \beta} k^\alpha k^\beta =
\kappa T_{\alpha \beta} k^\alpha k^\beta$ (where $T_{\alpha \beta}$
is the energy-momentum tensor). In the
comoving and synchronous coordinates, for pressure-less matter, $T_{\alpha \beta} k^\alpha k^\beta =
\kappa \rho k^0 k^0$. 
The Dyer--Roeder approach assumes homogeneity ($\sigma = 0 = \delta \rho$)
but takes into account that light propagates through vacuum. Therefore,
 $\Omega_m$ which photons `feel' is different than true $\Omega_m$. 
This is modelled by a constant parameter $\alpha$
(of value between 0 and 1) the multiplies  $\Omega_m$.
In this case (\ref{dsr}) reduces to 

\begin{equation}
\frac{{\rm d}^2 D_A}{{\rm d} z^2} + 
\left(\frac{1}{H}\frac{{\rm d} H}{ {\rm d} z} + \frac{2}{1+z} \right) \frac{{\rm d} D_A}{ {\rm d} z}
 + \frac{3\alpha  \Omega_m H_0^2}{2H^2}  (1+z)  D_A=0,
\label{ddre}
\end{equation}
where
$ H(z)=H_0 \sqrt{\Omega_{m} (1+z)^3+\Omega_{k}(1+z)^2 + \Omega_{\Lambda}}.$
The initial conditions needed to solve (\ref{ddre}) are: $D_A = 0$ and 
${{\rm d} D_A}/{ {\rm d} z} = 1/H_0$. 

Generalization to $\alpha(z)$ was explored by 
Linder (1988) who suggested several algebraic forms  like
$\alpha(z) = \alpha_0 + \alpha_1 z$ or
$\alpha(z) = \alpha_* + \alpha_2 (1+z)^\gamma$.
Santos \& Lima (2006) proposed
$\alpha(z) = \beta_0 (1+z)^{3 \gamma} / [ 1+ \beta_0 (1+z)^{3 \gamma}]$.
{This paper studies the following form of $\alpha(z)$}

\begin{equation}\label{alphaz}
\alpha(z) = 1 +  {\cal D}(z) \av{\delta}_{1D}
\end{equation}
{where $\av{\delta}_{1D}$ is the mean of present-day density fluctuations along the line of sight,
and ${\cal D}(z)$ describes its evolution. Below it is shown that if
${\cal D} = {(1+z)^{-5/4}}$, then the 
Dyer--Roeder equation gives results consistent with the results
obtain under the assumption of the lensing approximation.}

\section{Linear perturbations and the lensing approximation}\label{Lena}

Writing the distance as

\begin{equation}
 D_{A}(z)= \bar{D}_{A} ( 1 + \delta_D),
\label{dlen}
\end{equation}
where $\bar{D}_{A}$ the distance in the homogeneous universe, one can derive formula for $\delta_D$ using the linear perturbative scheme.
The most general form was presented and discussed by Pyne \& Birkinshaw (2004), Bonvin et al. (2006), Hui \& Greene (2006), {and Enqvist, Mattsson \& Rigopoulos (2009).}
Excluding the contribution from the motion of the observer and source, and taking the leading term, $\delta_D$ reduces to

\begin{equation}\label{dDBa}
\delta_D =
- \int\limits_0^{\chi_e} {\rm d} \chi
\frac{ \chi_e - \chi}{\chi_e} \chi \nabla^2 \phi(\chi),
\end{equation}
where $\chi$ is the comoving coordinate $d \chi = dz/H(z)$,
$\phi$ is the gravitational potential which can be related to density perturbations $\rho \delta$ via the  Poisson equation
$ \nabla^2 \phi = \frac{4 \pi G}{c^2} a^2 \rho \delta$.
Equation (\ref{dDBa})  is equivalent to the 
convergence in the lensing approximation and is known as the Born approximation.
As seen voids ($\delta <0$) increase the distance
while regions of $\delta >0$ decrease it.

In order to solve (\ref{dDBa}) one needs to know $\nabla^2 \phi$ along the line of sight. Using the Poisson equation, the gravitational potential is related to the density fluctuations.
Thus when calculating the variance, one can Fourier transform $\delta$ 
and use the matter power spectrum instead (Munshi \& Jain 2000, 2001).
However, in this paper we are not interested in the variance
or higher oder moments, just in the distance itself. Therefore, to solve (\ref{dDBa}) 
actual density fluctuations along the line of sight are needed.

The preset-day fluctuations are non-linear and therefore no longer Gaussian. Only when 
density fluctuations are in the linear regime, their 
probability distribution function (PDF) can be approximated by the Gaussian PDF.
Once density fluctuations are in the non-linear regime
they are no longer Gaussian (there is no symmetry as 
$-1 \leq \delta < \infty$).
However, it has been shown that in the non-linear regime
density fluctuations can  be approximated by  
the one-point log-normal PDF (Kayo, Taruya \& Suto 2001; Lahav \& Suto 2004)

\begin{equation}\label{nlPDF}
P(\delta) = \frac{1}{\sqrt{2 \pi \sigma_{nl}^2} } 
\exp \left[ - \frac{ (\ln(1+\delta) + \sigma_{nl}^2/2)^2 }{2 \sigma_{nl}^2} \right] \frac{1}{1+\delta},
\end{equation}

where
\begin{equation}\label{nlVar}
\sigma_{nl}^2 = \ln [ 1 + \sigma_R^2] \quad {\rm and} \quad \sigma_R^2 = \frac{1}{2\pi^2} \int\limits_0^\infty {\rm d} k {\cal P}(k) W^2(kR) k^2,
\end{equation}
where ${\cal P}(k)$ is the matter power spectrum.

Two methods for generating density fluctuations along the line of sight  {are} considered.
Both have a log-normal PDF of $\delta$ (3D). 
The first method ensures that $\av{\delta}_{1D} \approx 0$ (the mean of the
present-day fluctuations along the line of sight). The second method,
on the other hand, allows for $\av{\delta}_{1D} \ne 0$.

\begin{figure}
\includegraphics[scale=0.3]{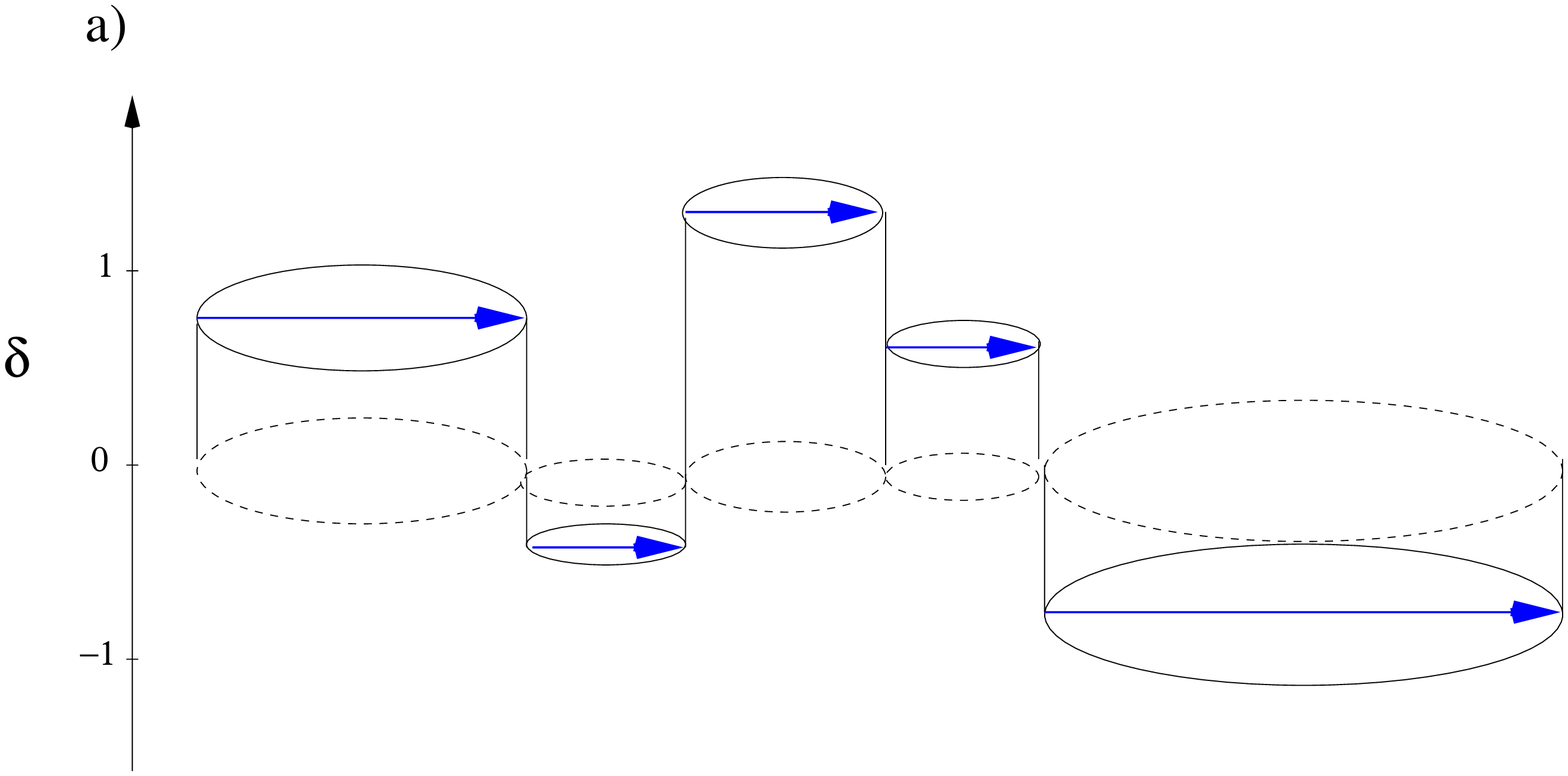}
\includegraphics[scale=0.17]{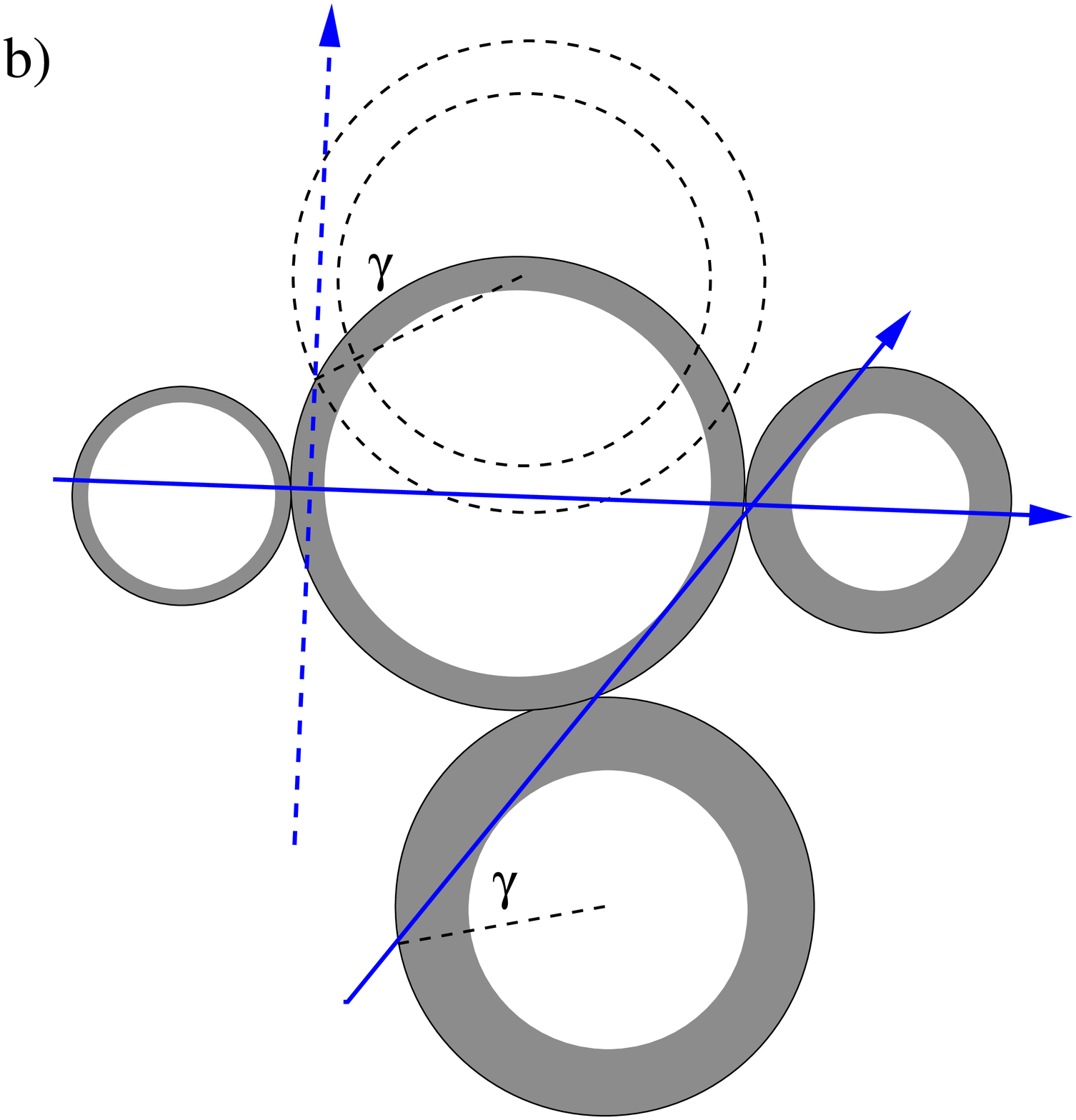}
\includegraphics[scale=0.3]{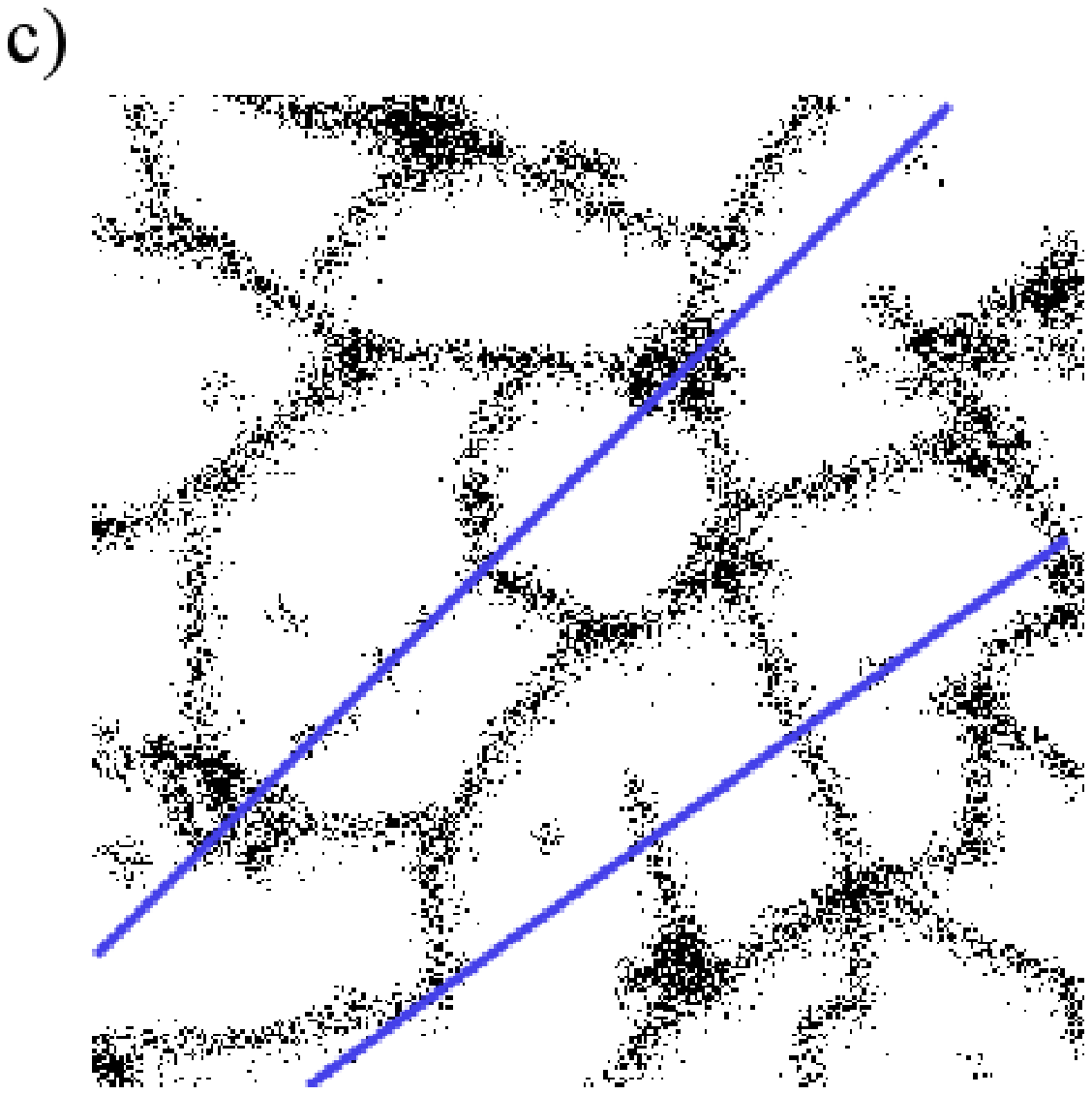}
\caption{Density inhomogeneities along the line of sight.
Panel a):  density fluctuations generated from the log-normal 
PDF (\ref{nlPDF}) -- when the light ray exits one structure, the next one is generated. Panel b): compensated density fluctuations -- when the light ray exits a structure, another one is generated together with an
initial angle at which the ray enters the structure. 
When $\gamma$ is large then the light enters and exits the structure at large angles.
In this case structures can overlap.
Panel c): light propagation in the real universe.} \label{fig1}
\end{figure}

\begin{figure}
\includegraphics[scale=0.6]{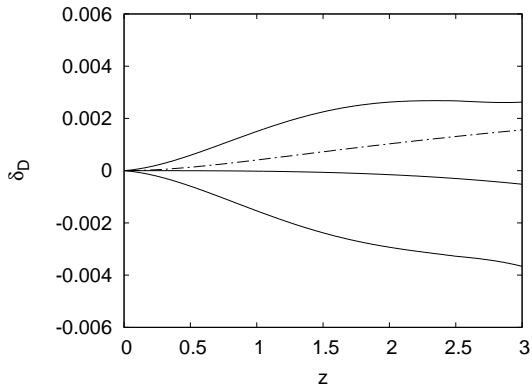}
\caption{Distance correction $\delta_D$ within the weak lensing
approximation for a model with a negligible mean of
density fluctuations along the line of sight (Sec. \ref{d1D0}).
Dash-dotted line presents $\delta_D$ obtained using the Dyer--Roeder relation with the smoothness parameter $\alpha =0.99$ ($\alpha =1$ is equivalent to the standard Friedmann relation).} \label{fig2}
\end{figure}

\begin{figure}
\includegraphics[scale=0.6]{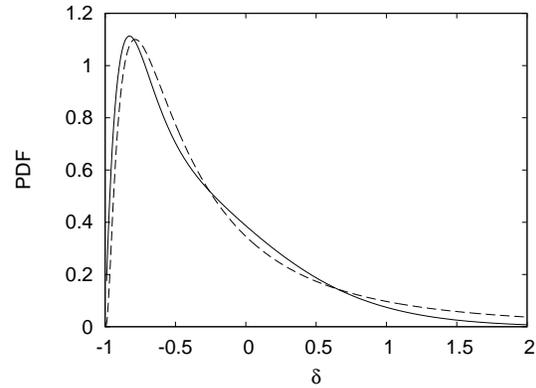}
\caption{PDF of $\delta$ smoothed in a sphere of radius 4 Mpc
for a model discussed in Sec. \ref{d1Dn0} (solid line)
and the log-normal PDF (\ref{nlPDF}) of $\delta$ smoothed also in a sphere of radius 4 Mpc (dashed line).} \label{fig3}
\end{figure}

\begin{figure}
\includegraphics[scale=0.6]{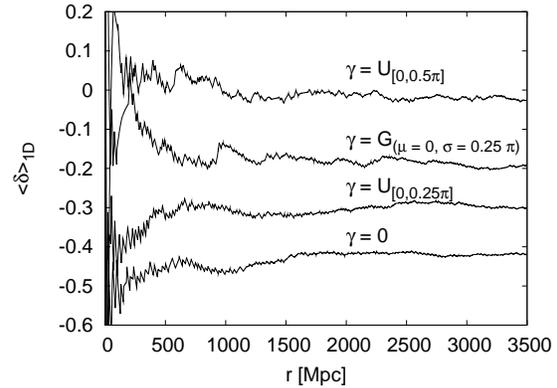}
\caption{Mean density fluctuations along the line of sight (\ref{md1d})
for a model discussed in Sec. \ref{d1Dn0}. The upper most curve presents a model with 
$\gamma$  generated from a uniform distribution 
between 0 and 0.5 $\pi$. The second curve presents a model with  
 $\gamma$ generated from the Gaussian distribution with the mean equal to zero
and the standard deviation equal to  $0.25 \pi$. In the third model
$\gamma$ is generated from a uniform distribution 
between 0 and $0.25 \pi$. 
The bottom curve presents a model with $\gamma =0$.} \label{fig4}
\end{figure}

\section{Results}\label{res}

\subsection{Vanishing mean of the density fluctuations along the line of sight}\label{d1D0}

First let us focus on generating the
density fluctuations using directly the log-normal PDF.
Using (\ref{nlPDF}) $\delta$ smoothed in a sphere of radius $R$
can be generated. Density fluctuations in this model
are schematically presented in Fig. \ref{fig1}a --
when the light ray exits one sphere it enters another one of
different $R$ and $\delta$ (for details see Appendix \ref{A1D0}).

The distance correction $\delta_D$ calculated from (\ref{dDBa})
is presented in Fig. \ref{fig2}. 
{One point about Fig. \ref{fig2}. needs to be emphasized. 
As follows from (\ref{dDBa}) the distance correction
depends on the position of the source,
for example $\delta_D(z=0.5)$ for a source at $z_* = 2$ is
different than $\delta_D(z=0.5)$ for a source at $z_* = 0.5$
even if density fluctuations along the line of sight are the same.
Thus, the distance correction presented in Fig. \ref{fig2} is the distance correction 
for the source at given $z$, i.e. $\delta_D(z=z_*)$.
Fig. \ref{fig2} presents the mean and variance:
at each $z$ (a discrete number of $z$ with an interval $\Delta z =0.01$ was considered)
based on 100,000 runs 
(each with a different distribution of density fluctuations along the line of sight)
the mean and variance were calculated. The mean and variance are
represented by solid lines.}
For comparison,
the Dyer--Roeder approximation with $\alpha = 0.99$ is also presented
(when $\alpha=1$ the Dyer--Roeder approximation reduces to the standard Friedmann
relation for the distance).
As seen $\delta_D$ is of negligible amplitude $\sim 10^{-3}$.
Thus, for this configuration,
and under considered assumptions, 
 there is no need to 
take into account inhomogeneities.
Also, another important fact
is that $\delta_D$ in the weak lensing approximation
is negative, while the Dyer--Roeder approximation implies positive
$\delta_D$. This is a consequence of 
positive density fluctuations along the line of sight.
{For example, for a source at $z=3$ (based on 100,000 runs) the mean
of the present-day density fluctuations along the line of sight
defined as}
\begin{equation}\label{md1d}
\av{\delta}_{1D}(r)  = \frac{1}{r} \int\limits_0^r {\rm d} \tilde{r} \delta(t_0,\tilde{r}),
\end{equation}
is $\av{\delta}_{1D} = 6 \times 10^{-3}$ and the standard deviation $2 \times 10^{-2}$,
which implies $\delta_D = - 5 \times 10^{-4} \pm 3 \times 10^{-3}$.
Thus, as pointed out by R\"as\"anen (2009)
in order to properly describe the effect of
clustering, the smoothness parameter should be larger than 1,
not less than 1 as in the Dyer--Roeder approximation.
However, in this case the difference as well as the total correction are negligible.
{However, even if the mean is negligible, the variance 
can provide additional information -- see Linder (2008) for a discussion.}

\subsection{Non-zero mean of the density fluctuations
along the line of sight}\label{d1Dn0}

In the previous section, random matter
fluctuations along the line of sight were considered. In such a
case $\av{\delta}_{3D}=0$ implies $\av{\delta}_{1D}=0$,
and the distance correction is negligible.
However, present-day matter fluctuations in the Universe
are not purely random but are organized -- matter in the Universe forms the cosmic web.
In this case
$\av{\delta}_{3D}=0$ does not necessarily imply 
$\av{\delta}_{1D}=0$.
As seen in Fig \ref{fig1}c the cosmic web 
contains  large voids with fairly compact filaments.
Therefore, photons spend more time in voids than
in overdense regions.
In order to model this phenomenon 
let us consider a universe that consists of voids
surround by filaments, each structure compensated.
Thus, by the construction a 3D average of density fluctuations is zero  
(if averaged over sufficiently large scale).
Moreover, if the parameters of the system are adjusted
(see Appendix \ref{A1Dn0} for details)
then the PDF of density fluctuations can be almost log-normal.
This feature is presented in  Fig. \ref{fig3},
which shows the PDF of density fluctuations smoothed within a sphere of 
radius 4 Mpc.
As seen the PDF is similar to log-normal PDF,
apart from high $\delta$ where PDF is of lower amplitude
-- but see Jain, Seljak \& White (2000) where 
the PDF of density fluctuations smoothed on scales of 3 $h^{-1}$ Mpc
has also a lower amplitude for high $\delta$ than the log-normal distribution.

When the light ray exits one compensated structure
before it enters another one, not only parameters of the next
structure are generated, but also an angle $\gamma$, at which the light ray 
enters another structure. Therefore, 
4 different methods are going to be considered

\begin{enumerate}
\item $\gamma = 0$, the light ray always
enters another structure at $\gamma = 0$, thus it passes through
the centre of the structure.

\item $\gamma$  
is randomly generated from a uniform distribution 
between 0 and 0.5 $\pi$. 
However,  as shown in Fig. \ref{fig1}b in this case
structures can overlap.

\item to reduce overlapping, $\gamma$ 
is generated from the Gaussian distribution with
the mean 0 and $\sigma = 0.25 \pi$
(if $|\gamma| > 0.5 \pi$ then $|\gamma| \to \pi - |\gamma|$).

\item $\gamma$ is randomly generated from a uniform distribution 
between 0 and 0.25 $\pi$. 

\end{enumerate}

\begin{figure*}
\includegraphics[scale=0.65]{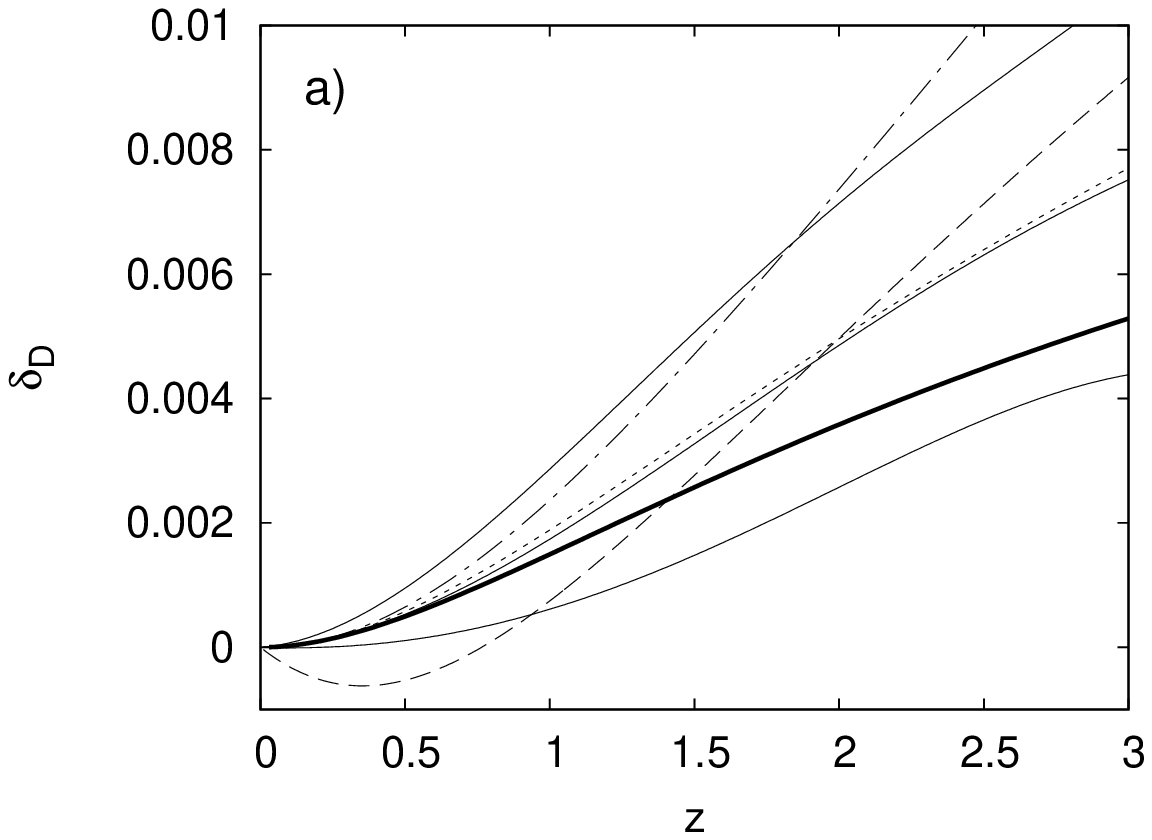}
\includegraphics[scale=0.65]{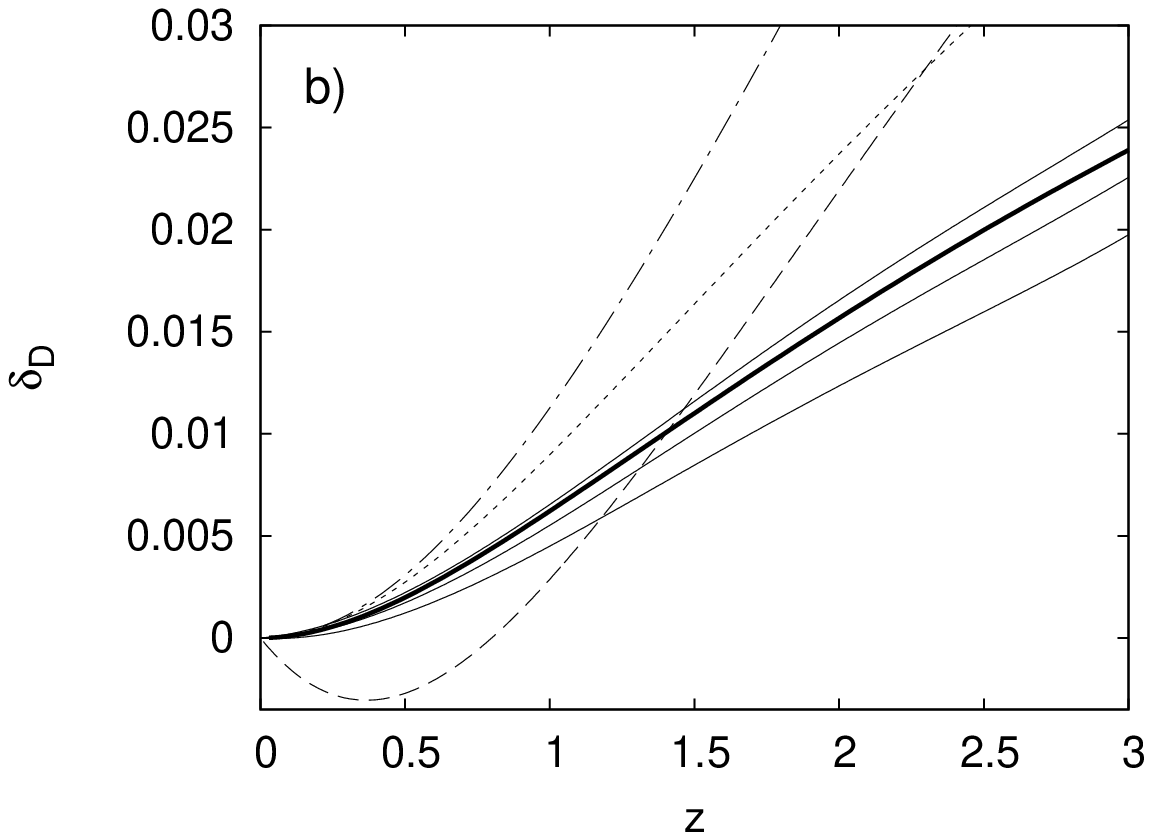}
\includegraphics[scale=0.65]{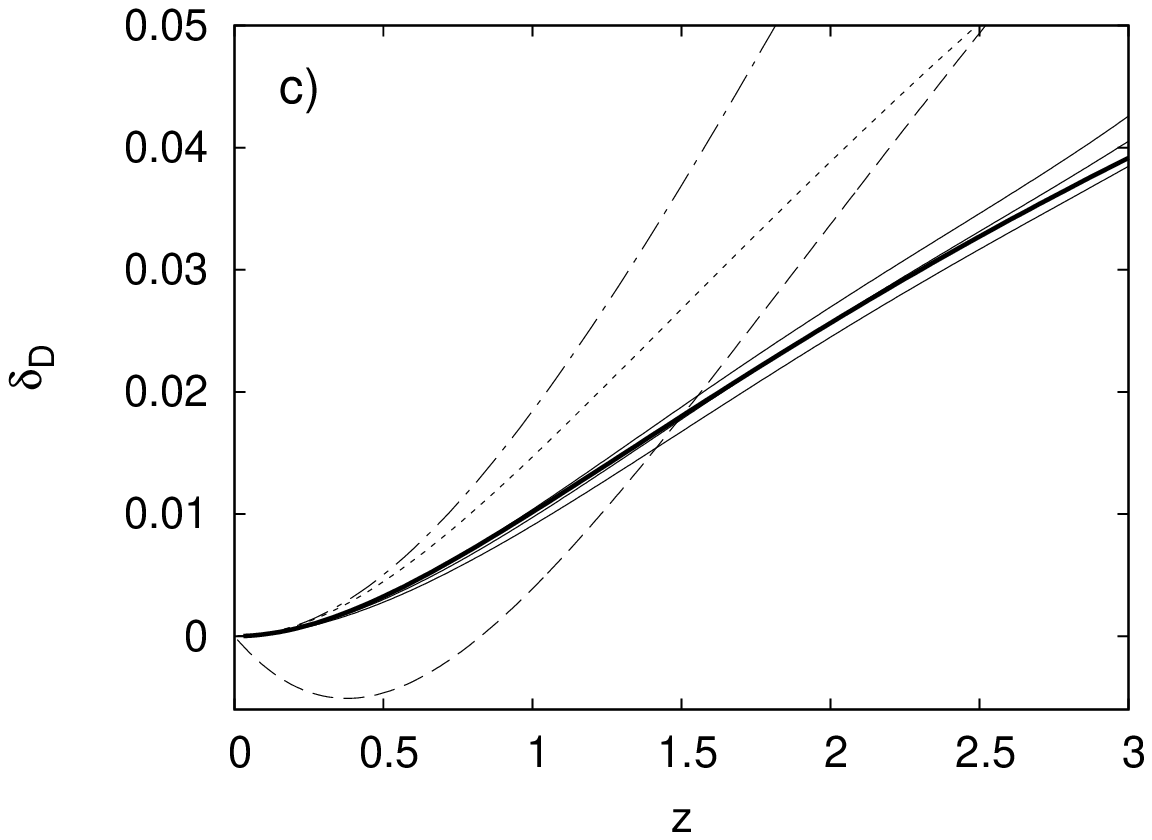}
\includegraphics[scale=0.65]{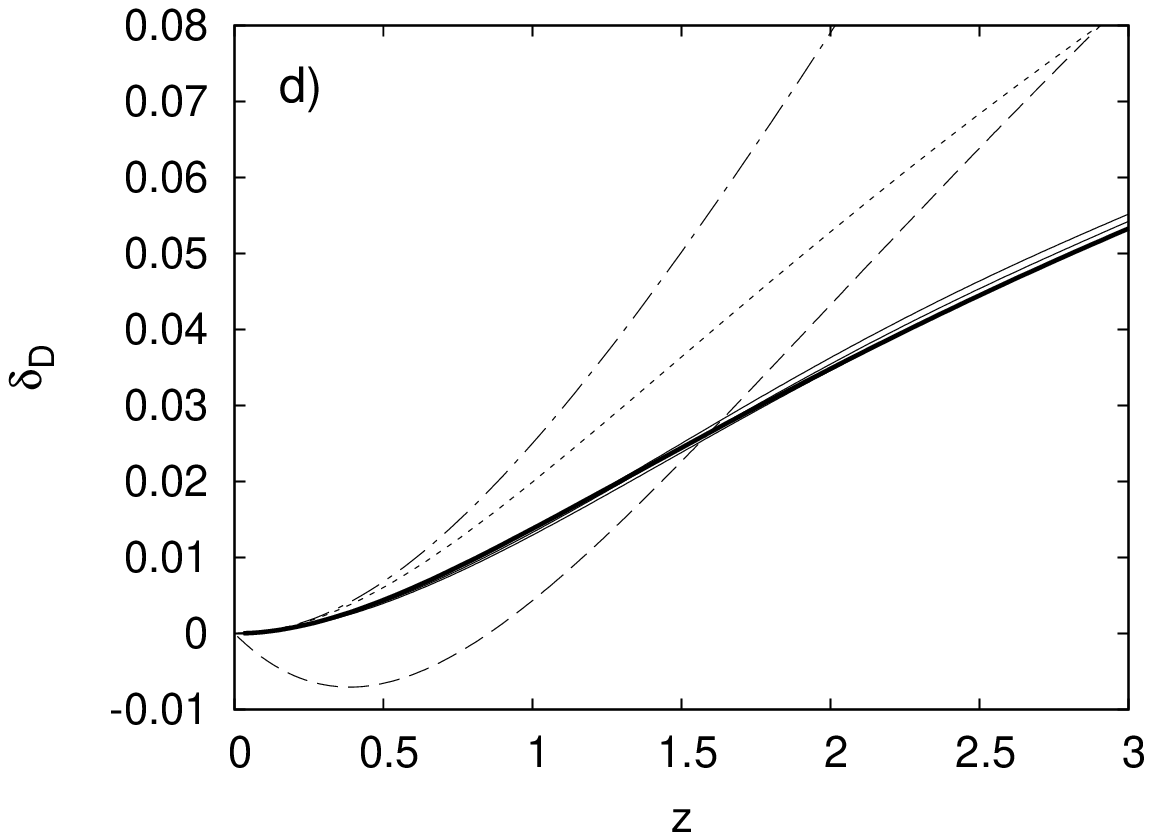}
\caption{Distance correction $\delta_D$ for a model
 discussed in Sec. \ref{d1Dn0}. 
The weak lensing approximation (\ref{dDBa})
is presented by {thin solid curves (mean + variance), the dash-dotted} lines
present the Dyer--Roeder approximation with a constant smoothness parameter
$\alpha = 1 + \av{\delta}_{1D}$, 
the dotted lines represent models with $\alpha(z) = 1+ \delta(z)$,
where $\delta(z)$ is given by (\ref{deltaev}),
{the dashed lines
 present the distance correction for models with a
perturbed expansion rate (see text for details)},
and the {thick solid lines represent the modified} Dyer--Roeder relations
with  $\alpha = 1 + \av{\delta}_{1D}/(1+z)^{5/4}$.
Panel a): $\gamma$  generated from a uniform distribution 
between 0 and 0.5 $\pi$, $\av{\delta}_{1D} \approx -0.04$. Panel b): 
$\gamma$ generated from the Gaussian distribution with
the mean 0 and $\sigma = 0.25 \pi$, $\av{\delta}_{1D} \approx -0.19$.
Panel c): $\gamma$ generated from a uniform distribution 
between 0 and 0.25 $\pi$ $\av{\delta}_{1D} \approx -0.31$. Panel d): $\gamma = 0$, $\av{\delta}_{1D} \approx -0.42$.} \label{fig5}
\end{figure*}

The distance correction  $\delta_D$ (the mean and variance) is presented in Fig. \ref{fig5}.
As seen if $\gamma$ is randomly generated from a uniform distribution 
between 0 and 0.5 $\pi$ then the mean of density fluctuations
along the line of sight is small and $\delta_D$ is almost negligible
(cf. Brouzakis, Tetradis \& Tzavara 2008; Vanderveld, Flanagan \& Wasserman 2008).

However, as pointed out above, if $\gamma$ changes between $0$ and $0.5\pi$ 
then structures can overlap (see Fig \ref{fig1}b).
To reduce  the overlapping, $\gamma$ needs to by chosen from a smaller range of angles.
In this case $\delta_D$ increases.
This is because the light rays propagate more likely through voids than filaments.

In Fig. \ref{fig5} the Dyer--Roeder relation
is also presented. The {dash-dotted} line presents
$\delta_D$ obtained from the original Dyer--Roeder formula,
the dashed line and dotted line prenent the evolving smoothness parameter (\ref{alphaz}).
{Intuitively, if $\av{\delta}_{1D}$ were just an ordinary density perturbation,
in the linear regime, it should evolve according to  (\ref{deltaev}).
Thus, the dotted line presents a model with $\alpha(z) = 1+ \delta(z)$,
where $\delta(z)$ is given by (\ref{deltaev}) (with an initial condition 
$\delta(z=0) = \av{\delta}_{1D}$).

Following Mattsson (2010) let us consider the
Dyer-Roeder relation with a perturbed expansion rate.
From the continuity equation 
$\dot{\rho} + 3 H \rho = 0$, the perturbation in the
expansion rate are $\Delta H = -\dot{\delta} / 3$
(where $\dot{\delta}$ follows from (\ref{deltaev})
with an initial condition $\delta(z=0) = \av{\delta}_{1D}$).
We use this relation to calculate the perturbed $H(z)$ and ${\rm d}H/{\rm d}z$,
and we then insert it to (\ref{ddre}), and using 
$\alpha(z) = 1+ \delta(z)$,
where $\delta(z)$ is given by (\ref{deltaev}), we get the distance
correction that is presented using the dashed lines.

As seen, neither a constant $\alpha$, nor the evolving one
(with $\delta(z)$ given by (\ref{deltaev})),
or with a pertubed expanion rate, are 
consistent
with the weak lensing approximation (expect for the case where $\delta_D$
is almost negligible). However, 
if $\alpha(z) = 1+ \av{\delta}_{1D}/(1+z)^{\gamma}$, where $\gamma = 5/4$
then the Dyer--Roeder approximation
produces results comparable with the weak lensing approximation.
As seen, the evolution of $\alpha$ is not as intuitively expected,
i.e. it does not directly follow from (\ref{deltaev}).
Here an empirical approach was employed, and 
it was found that if $\gamma \approx 5/4$
then the modified, in this way, Dyer-Roeader relation 
leads to the agreement with the lensing approximation.
}

An important result of the above analysis is that a non-zero mean of density fluctuations along the line of sight
can modify the distance by a few percent.
As cosmological observations are now reaching the precision
of a few percent, a proper handling of the distance
is essential.
Otherwise the errors in the estimation of the distance can become a major source of systematics.
For example a proper handling of distance will be of great importance 
in the future analysis of BAO experiments (Bolejko 2010).

\section{Conclusions}\label{dis}

The distance-redshift relation plays a central role in  
cosmology. 
If the homogeneous Friedmann model correctly describes the evolution of the  universe
on large scales and, in addition, if density fluctuations
along the line of sight vanish after averaging, then
within the linear approximation, the distance correction 
is negligible -- it is sufficient to apply the Friedmann relation.
However, if only the mean of density fluctuations along
the line of sight is not zero then,
even within the linear approximation, 
 the distance 
can change by several percent.

In Sec. \ref{d1Dn0} it was shown that a vanishing 3D average of
density fluctuations does not imply that the mean of density fluctuations
along the line of sight is zero.
It is argued that in the real universe this may be the case. 
In the real universe voids occupy large regions 
while overdensities are more compact. Moreover, if light
propagates for a long time through filaments then
it is more likely to be absorbed or scattered.
Thus, if a remote galaxy is observed then 
most likely its photons propagated through emptier rather then denser regions.
{In this case 
the weak lensing approximation produces results that
are similar to results obtained from the modified version of the Dyer--Roeder
equation.} The modified relation is

\begin{eqnarray}\label{mdre}
&& \frac{{\rm d}^2 D_A}{{\rm d} z^2} + 
\left(\frac{1}{H}\frac{{\rm d} H}{ {\rm d} z} + \frac{2}{1+z} \right) \frac{{\rm d} D_A}{ {\rm d} z} + \nonumber \\
 && \frac{3}{2} \frac{H_0^2}{H^2} \Omega_m (1+z)
\left(1 +  \frac{\av{\delta}_{1D}}{(1+z)^{5/4}} \right)
 D_A=0,
\end{eqnarray}
(with initial conditions $D_A = 0$ and  ${{\rm d} D_A}/{ {\rm d} z} = 1/H_0$).
Thus apart from the background cosmological parameters, the only free parameter left
is the mean of density  fluctuations along the line of sight $\av{\delta}_{1D}$.
Since (\ref{mdre}) is just an ordinary differential equation, 
it is as easy to implement it and solve numerically as the standard
relation for the distance in the Friedmann model.
The mean of density fluctuations along the line of sight $\av{\delta}_{1D}$
could either be deducted from the galaxy redshift surveys or N-Body simulations.

{ 
It should be noted that the analysis presented here was 
based on the linear approximation (the lensing approximation), i.e. higher order corrections to the distance 
were not taken into account. It is likely that quadratic corrections 
may lead to non-negligible $\delta_D$ even when $\av{\delta}_{1D} \approx 0$.
In this case the presented above modified version of the Dyer--Roeder
equation would not be consistent with the actual non-linear distance-redshift relation,
as when $\av{\delta}_{1D} = 0$ it reduces to the Friedmannian relation.}.

The most important conclusion of this paper is
that even within the idealized case of large scale 
homogeneous universe (with only Mpc-scale inhomogeneities) the distance can be different
than in the homogeneous universe. 
The difference can be of order of a few percent.
Thus, with an increasing precision of cosmological observations 
an accurate estimation of the distance is essential. Otherwise errors 
due to miscalculation the distance can become a major source of systematics.

\section*{Acknowledgments}

I would like to thank Eric Linder,
Teppo Mattsson, Syksy R\"as\"anen,
and the referee for useful discussions and suggestions.
The support of the Go8 Fellowship and the hospitality of the ANU
Centre for Gravitational Physics is also acknowledged.

\appendix

\section{Algorithm of calculating $\delta_D$ 
for model discussed in Sec. 4.1}\label{A1D0}

The algorithm of calculating $\delta_D$ along the line of sight 
is as follows

\begin{enumerate}
\item
The radius of a structure $R$ is randomly generated from a uniform distribution,
from 0 to 3 Mpc.
\item
Then $\sigma_R$ is calculated using  (\ref{nlVar}).
{ 
The primordial power spectrum was chosen in agreement with the WMAP7 data (Komatsu et al. 2010):
$A k^n_s$ with $n_s = 0.969$ and the amplitude $A$ chosen so that $\sigma_8 = 0.803$.
The transfer function was calculated according to Eisenstein \& Hu (1998).}

\item
An initial value of a density fluctuation $\delta_0$ is generated
from the log-normal distribution (\ref{nlPDF}).

\item
The evolution of $\delta$ (at a fixed point) is
calculated using the linear approximations  (Peebles 1980)
\begin{equation}\label{deltaev}
\ddot{\delta} + 2 \frac{\dot{a}}{a} \dot{\delta} = \frac{4 \pi G}{c^2} \rho \delta.
\end{equation}
\item
Using the Poisson equation, $\nabla^2 \phi$ is calculated and inserted  into (\ref{dDBa})
which is solved from $\chi_i$ to $\chi = \chi_i + \chi(2R)$.

\item
Steps (i)--(v) are repeated so  (\ref{dDBa}) is solved from $\chi = 0$ to $\chi_e$.
\end{enumerate}

The cosmological parameters are the same as derived
from the 7-year WMAP data: $H_0 = 71.4$ km s$^{-1}$ Mpc$^{-1}$, $\Omega_m = 0.262$,
$\Omega_\Lambda = 0.738$ (Komatsu et al. 2010).

\section{Algorithm of calculating $\delta_D$ for model discussed in Sec. 4.2}\label{A1Dn0}

The algorithm of calculating $\delta_D$ is as follows

\begin{enumerate}
\item
First the radius of a void $R_v$ is  generated from the Gaussian distribution 
with the mean of 12 Mpc and the standard deviation of 2 Mpc.
\item Density within the void $\Omega_v$ is generated from the Gaussian distribution with the mean 
 $0.2 \Omega_m$ and $\sigma = 0.27 \Omega_m$.
If a generated in this way $\Omega_v$ is lower than $0.01 \Omega_m$
then the generation is repeated.
If after 6 times it is still less than $0.01 \Omega_m$
then $\Omega_v$ is generated for a uniform distribution between 
 $0$ and $0.01 \Omega_m$. 
If $\Omega_v \geq \Omega_m$
then $\Omega_v$ is generated one more time.
If after 6 times $\Omega_v \geq \Omega_m$ then
its value is chosen for a uniform distribution 
from $0.85 \Omega_m$ to $\Omega_m$.
\item Density of the surrounding shell 
$\Omega_s$ is generated from the Gaussian distribution with
the mean of $1.75 \Omega_m$ and $\sigma = 0.7 \Omega_m$.
If $\Omega_s \leq \Omega_m$ then its value
is generated again. If after 6 times 
 $\Omega_s \leq \Omega_m$ then its value is generated for a uniform distribution
between $1.75 \Omega_m$ and $1.95  \Omega_m$.
\item
The condition, that the  structure is compensated implies that the radius of the 
whole structure is

\[R = R_v \left( \frac{\Omega_s - \Omega_v}{\Omega_s - \Omega_m} \right)^{1/3}.\]
\item The angle at which the light ray enters  the structure is generated
using 4 different methods  -- for details see Sec. \ref{d1Dn0}.
\item The evolution of $\delta$ (at a fixed point) was 
calculated using (\ref{deltaev}).
\item 
Integral (\ref{dDBa})
is solved from $\chi_i$ to $\chi_f$ (where $\chi_i$ is the comoving
coordinate of the entry point and $\chi_f$ the point where
the light ray exits the structure).
\item 
Steps (i)--(vii) are repeated so  (\ref{dDBa}) is solved from $\chi = 0$ to $\chi_e$. 
\end{enumerate}

If instead of the Gaussian,
the uniform PDF is used (steps (i)--(iii)) then the PDF of $\delta$
is  not log-normal PDF as presented in Fig. \ref{fig3}.

The cosmological parameters are the same as derived
from the 7-year WMAP data: $H_0 = 71.4$ km s$^{-1}$ Mpc$^{-1}$, $\Omega_m = 0.262$,
$\Omega_\Lambda = 0.738$ (Komatsu et al. 2010).


\bsp

\label{lastpage}

\end{document}